\documentclass[conference]{IEEEtran}
\IEEEoverridecommandlockouts

\usepackage{cite}
\usepackage{amsmath,amssymb,amsfonts}
\usepackage{algorithmic}
\usepackage{graphicx}
\usepackage{textcomp}
\usepackage{tabularx}
\usepackage{xcolor}
\usepackage{url} 
\usepackage[colorlinks=true, linkcolor=blue, urlcolor=blue, citecolor=blue]{hyperref}

\def\BibTeX{{\rm B\kern-.05em{\sc i\kern-.025em b}\kern-.08em
    T\kern-.1667em\lower.7ex\hbox{E}\kern-.125emX}}
\begin{document}

\title{NeuroIncept Decoder for High-Fidelity Speech Reconstruction from Neural Activity\\
\thanks{© 20XX IEEE. Personal use of this material is permitted. Permission from IEEE must be obtained for all other uses, in any current or future media, including reprinting/republishing this material for advertising or promotional purposes, creating new collective works, for resale or redistribution to servers or lists, or reuse of any copyrighted component of this work in other works.}
\thanks{This work was supported by grant PID2022-141378OB-C22 funded by MICIU/AEI/10.13039/501100011033 and by ERDF/EU.}
}

\author{\IEEEauthorblockN{1\textsuperscript{st} Owais Mujtaba Khanday}
\IEEEauthorblockA{\textit{CITIC-UGR}\\
\textit{University of Granada}\\
Granada, 18071, Spain \\
owaismujtaba@ugr.es}
\and
\IEEEauthorblockN{2\textsuperscript{nd} José L. Pérez-Córdoba}
\IEEEauthorblockA{\textit{CITIC-UGR}\\
\textit{University of Granada}\\
Granada, 18071, Spain \\
jlpc@ugr.es}
\and
\IEEEauthorblockN{3\textsuperscript{rd} Mohd Yaqub Mir}
\IEEEauthorblockA{\textit{Epilepsy Centre, Dept. Clinical Sciences}\\ 
\textit{Lund University Hospital}\\  
Lund, 221 00, Sweden\\ 
yaqubmir@gmail.com}

\and
\IEEEauthorblockN{4\textsuperscript{th} Ashfaq Ahmad Najar}
\IEEEauthorblockA{\textit{School of Computing Science and Engineering} \\
\textit{VIT Bhopal University}\\
Bhopal, India \\
ashfaqahmadnajar@vitbhopal.ac.in}

\and
\IEEEauthorblockN{5\textsuperscript{th} Jose A. Gonzalez-Lopez}
\IEEEauthorblockA{\textit{CITIC-UGR} \\
\textit{University of Granada}\\
Granada, 18071, Spain \\
joseangl@ugr.es}
}

\maketitle

\begin{abstract} 
    This paper introduces a novel algorithm designed for speech synthesis from neural activity recordings obtained using invasive electroencephalography (EEG) techniques. The proposed system offers a promising communication solution for individuals with severe speech impairments. Central to our approach is the integration of time-frequency features in the high-gamma band computed from EEG recordings with an advanced NeuroIncept Decoder architecture. This neural network architecture combines Convolutional Neural Networks (CNNs) and Gated Recurrent Units (GRUs) to reconstruct audio spectrograms from neural patterns. Our model demonstrates robust mean correlation coefficients between predicted and actual spectrograms, though inter-subject variability indicates distinct neural processing mechanisms among participants. Overall, our study highlights the potential of neural decoding techniques to restore communicative abilities in individuals with speech disorders and paves the way for future advancements in brain-computer interface technologies.
\end{abstract}

\section{Introduction}
    Speech disorders, resulting from damage to muscles, nerves, or vocal structures, impair sound production and can severely impact communication and quality of life \cite{jullien2021screening, Gonzalez2020}. These conditions, including stuttering and apraxia, are prevalent worldwide, especially among children, with 7.7\% of US children aged 3-17 and 11\% of those aged 3-6 experiencing speech-related disorders annually \cite{bib1, bib2, bib3}. Neuroprostheses (also known as brain-computer interfaces), which interface with the nervous system to restore lost functions, offer innovative solutions for communication impairments, particularly in conditions like amyotrophic lateral sclerosis (ALS), where speech muscles are affected but cognition remains intact \cite{Gonzalez2020, neuro1, neuro2, neuro3}. These devices, whether implanted or external, enable more natural and effective interactions than traditional communication aids, providing critical support for individuals with severe motor limitations \cite{bib4, bib5}.

    Automatic speech recognition (ASR) and speech synthesis are the two main approaches for decoding speech from neural signals. The ASR-based approach \cite{asr1, asr2, asr3}  utilizes specialized software and models to convert neural signals into textual representations, while the speech synthesis approach aims to generate audible audio directly from the neural signals. In this work, we focus on this second approach. Traditional approaches for decoding and synthesizing speech spectrograms from neural signals, such as linear models and formant synthesis with Kalman filters, achieved limited performance with low-quality synthesized audio and Pearson correlation coefficients (PCC) below 0.7 \cite{bib6, bib7, bib8}. Recent advances leverage deep neural networks (DNNs), including convolutional and recurrent architectures, which improve intermediate representations and quality of speech synthesis \cite{bib9, bib10, bib11}. However, challenges remain due to the noisy and redundant nature of neural signals, low signal-to-noise ratios (SNRs), and temporal misalignments between neural and speech signals, making it difficult to accurately capture speech-related patterns like prosody and articulation \cite{asr2, a3}. Thus, speech synthesis from neural signals remains an open problem.
    

    In this study, we introduce a novel technique for decoding audible speech directly from neural signals. The main novelty of our work is the development of a novel DNN architecture, NeuroIncept, which leverages a multi-scale feature extraction pipeline through the incorporation of Inception modules. As outlined in Section \ref{ssec:neuroincept}, this architecture enables the model to capture a wide range of temporal and spectral patterns, facilitating the analysis of both fine-grained details and broader trends in neural data. This design addresses a critical limitation in prior approaches, such as \cite{bib9}, which rely on uniform filter sizes and may overlook the multiscale patterns crucial for precise decoding. Furthermore, our NeuroIncept decoder integrates temporal modeling via gated recurrent units (GRUs), ensuring robust handling of neural signal misalignments while effectively capturing temporal dependencies. By combining Inception modules with recurrent mechanisms, NeuroIncept delivers a comprehensive and adaptive feature representation, offering significant advancements over traditional methods.
    

\section{Methodology}
    \subsection{Dataset Description} \label{ssec:dataset}
        The dataset used in this study is publicly available \cite{verwoert2022dataset} and consists of stereotactic EEG (sEEG) recordings from 10 Dutch participants (5 Male and 5 Female; average age: 32 years) with pharmacoresistant epilepsy. Depth sEEG electrodes were implanted in the participants as part of their clinical treatment. The placement of the electrode was determined solely based on clinical requirements, primarily targeting the superior temporal sulcus, the hippocampus, and the inferior parietal gyrus. As a result, the number and locations of the electrodes varied between the participants. sEEG signals were recorded at either 2048 Hz or 1024 Hz, synchronized with participants’ speech ($F_s= 48$ kHz), while they read aloud a list of 100 words from the Dutch IFA corpus \cite{ifaCorpus}. The sEEG recordings were subsequently down-sampled to 1024 Hz, while speech signals were down-sampled to 16 kHz for further analysis. To ensure participant anonymity, pitch modulation of the audio recordings was applied using the LibROSA library \cite{librosa}.   
    
    \subsection{Signal Processing}
        \begin{figure}[t]  
            \centering 
            \footnotesize
            \includegraphics[width= \linewidth]{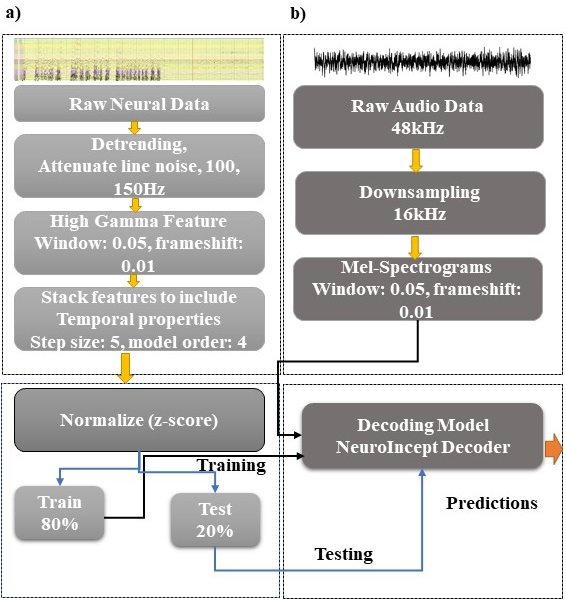} 
            \caption{Preprocessing pipeline for the sEEG and audio signals.} 
            \label{fig:dataprocessing}  
        \end{figure}
    
        The sEEG signals for each participant were parameterized as time-frequency features extracted from the high-gamma band (70-170 Hz), as shown in Figure \ref{fig:dataprocessing}a. This band was chosen because previous studies have shown that it contains information related to speech and language production and perception \cite{a3,band1, band2}. The raw sEEG data was first detrended to remove linear trends. A bandpass filter (70-170 Hz) was then applied to isolate the high-gamma-frequency components, while a notch filter targets line noise (50 Hz) and its two first harmonics (100 Hz, 150 Hz), further refining the signal. After filtering, the Hilbert transform was utilized to compute the analytic signal, which enables the extraction of the signal envelope capturing amplitude fluctuations within the high-gamma band. The processed sEEG signals underwent segmentation into overlapping temporal windows of 0.05s, with a frame-shift of 0.01s. Within each window, the mean amplitude was computed, producing a feature matrix for subsequent analysis. To incorporate temporal dynamics, each 0.05s window was further expanded by integrating the features from both the current and neighboring time windows. This process was achieved using a sliding window approach with a model order of 4 and step size of 5, enhancing temporal resolution and capturing dependencies across multiple time intervals. 
   
        The audio signals, on the other hand, were converted into logarithmic Mel-scaled spectrograms (logMel) with 128 spectral bins, as illustrated in Figure \ref{fig:dataprocessing}b. The logMel spectrograms were extracted from 0.05 s overlapping windows with a frame shift of 0.01 s using a Hanning window. The neural data for each participant was then normalized by z-score standardization, which enhances the comparability between data points and optimizes the subsequent model training procedures. 
    
    \subsection{Decoding Model Architecture} \label{ssec:neuroincept}
        \begin{figure}[t]  
            \centering  
            \footnotesize
            \includegraphics[width= \linewidth]{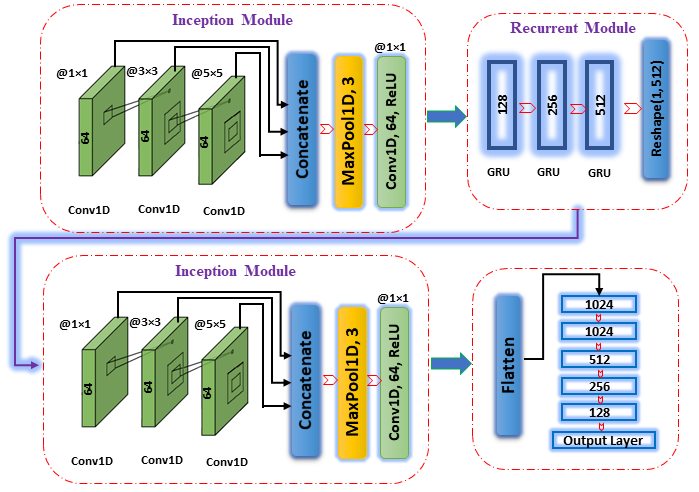}  
            \caption{NeuroIncept Decoder model architecture.} 
            \label{fig:model}  
        \end{figure}
        
        The NeuroIncept Decoder architecture shown in Figure \ref{fig:model}, is designed to efficiently process and analyze sequential data by combining the complementary strengths of Convolutional Neural Networks (CNNs) and Gated Recurrent Units (GRUs). Central to this architecture are two distinct, yet synergistic, modules: an Inception module, which serves as the primary feature extractor, and a Recurrent module, which is responsible for temporal pattern recognition.

        \textbf{Inception Module}: The Inception Module \cite{inception} of the NeuroIncept Decoder architecture acts as the primary feature extractor, adeptly processing input sequence data through multiple convolutional filters with varying kernel sizes: 1x1, 3x3 and 5x5. Each filter serves a unique role: the 1x1 convolution reduces the dimensionality of the data, preserving essential spatial information while streamlining computation; the 3x3 and 5x5 convolutions capture medium- and large-scale patterns, respectively, from the input sequences. The outputs from these operations are concatenated followed by the MaxPooling operation and the 1x1 convolutional layer to further reduce the spatial dimensions while integrating the pooled features. This approach allows the Inception Module to capture diverse temporal/spectral patterns from the sequence, thus enabling the model to process fine-grained and broader features in the neural data. In contrast, approaches such as \cite{bib9, bib10} use uniform filter sizes, which may fail to detect multiscale patterns critical for accurate decoding.
    
        \textbf{Recurrent Module} This module leverages GRU-based recurrent neural networks to capture temporal dependencies \cite{gru}. The first GRU layer, consisting of 128 units, processes the extracted features and returns a complete sequence of hidden states. This ensures that the succeeding GRU layers can operate with a full temporal context. As the data flows through successive GRU layers, the model progressively learns more intricate temporal patterns. The final GRU layer, with 512 units, produces a single output that encapsulates the entire sequence into a condensed summary. The Reshape layer then reformat this output into a tensor with a one time-step and 512 features, preparing the data for subsequent processing.

        The Recurrent Module output is passed through the second Inception module, then flattened into a one-dimensional vector. This vector is sent through several dense layers, gradually reducing sizes from 1024 to 128 units. These layers are designed to refine the extracted features and yield the model’s ultimate output. The implementation of our NeuroIncept Decoder architecture can be found at \url{https://github.com/owaismujtaba/NeuroInceptDecoder}.
    \subsection{Evaluation}
        The dataset for each participant, consisting of 5 minutes of simultaneous sEEG and audio signals while the participant read aloud a list of 100 Dutch words, was divided into training and validation sets in an 80-20\% ratio. In the training stage, the parameters of NeuroIncept architecture were optimized using the Mean Squared Error (MSE) loss function. To address the potential for overfitting, early stopping was implemented with a patience of 5 epochs.
        For evaluation, 1,000 samples were selected from the 20\% validation set for each participant using 10-fold cross-validation. The performance of the proposed method was evaluated using the Pearson Correlation Coefficient (PCC) to quantify the similarity between the spectrograms generated by the NeuroIncept Decoder and the original audio spectrograms. Additionally, the Spectral Temporal Glimpsing Index (STGI) \cite{edraki2022spectro} was employed to assess how effectively the predicted spectrograms retained the temporal and spectral characteristics of the original audio signals. As this study is primarily focused on accurately decoding neural signals into log-Mel spectrogram representations of audio, subjective listening tests were not conducted. Future research will address this limitation by incorporating vocoder systems to reconstruct audio waveforms from the decoded spectrograms, enabling more comprehensive assessments of neural-to-audio decoding performance.
    
\section{Results}
    In this section, we present the results achieved by the proposed NeuroIncept Decoder model in the speech synthesis task outlined above. First, in Section \ref{ssec_detailed_results}, we present the speech quality metrics obtained by our system for each subject. Then, in Section \ref{ssec:comparison_results}, we perform a quantitative and qualitative comparative analysis with other models from the literature.
    
    \subsection{Detailed Results} \label{ssec_detailed_results}
        \begin{table}[t]
            \centering
            \footnotesize
            \caption{Performance metrics on individual subjects.}
            \begin{tabular}{|c|c|cc|cc|}
            \hline
            \textbf{Participant} & \textbf{MSE} & \multicolumn{2}{c|}{\textbf{PCC}} & \multicolumn{2}{c|}{\textbf{STGI}} \\ \hline
            & & \textbf{Value} & \textbf{STD} & \textbf{Value} & \textbf{STD} \\ \hline
            sub-01 & 0.445 & 0.921 & 0.003 & 0.511 & 0.004 \\ \hline
            sub-02 & 0.511 & 0.926 & 0.002 & 0.477 & 0.005 \\ \hline
            sub-03 & 0.506 & 0.925 & 0.002 & 0.502 & 0.005 \\ \hline
            sub-04 & 0.522 & 0.938 & 0.004 & 0.479 & 0.005 \\ \hline
            sub-05 & 0.594 & 0.932 & 0.003 & 0.502 & 0.003 \\ \hline
            sub-06 & 0.409 & 0.944 & 0.002 & 0.552 & 0.004 \\ \hline
            sub-07 & 0.788 & 0.942 & 0.004 & 0.511 & 0.006 \\ \hline
            sub-08 & 0.652 & 0.897 & 0.005 & 0.526 & 0.005 \\ \hline
            sub-09 & 0.400 & 0.917 & 0.002 & 0.459 & 0.004 \\ \hline
            sub-10 & 0.498 & 0.838 & 0.007 & 0.522 & 0.004 \\ \hline
            \end{tabular}
            \label{tab:participants}
        \end{table}

    Table \ref{tab:participants} shows the average and standard deviation (std) for each metric obtained by our model among the ten participants in the database described in Section \ref{ssec:dataset}. The MSE values range from 0.400 to 0.788, reflecting differences in prediction accuracy. Higher correlation values suggest strong relationships and a high likelihood of accurate audio reconstruction from neural data. The small standard deviations of the correlations indicate consistent performance between participants. The STGI values range from 0.502 for sub-09 to 0.552 for sub-06, with minimal standard deviations, demonstrating stable individual scores.

    \begin{figure}[t]  
        \centering 
        \footnotesize
        \vspace{-0.3cm}
        \includegraphics[width= \linewidth]{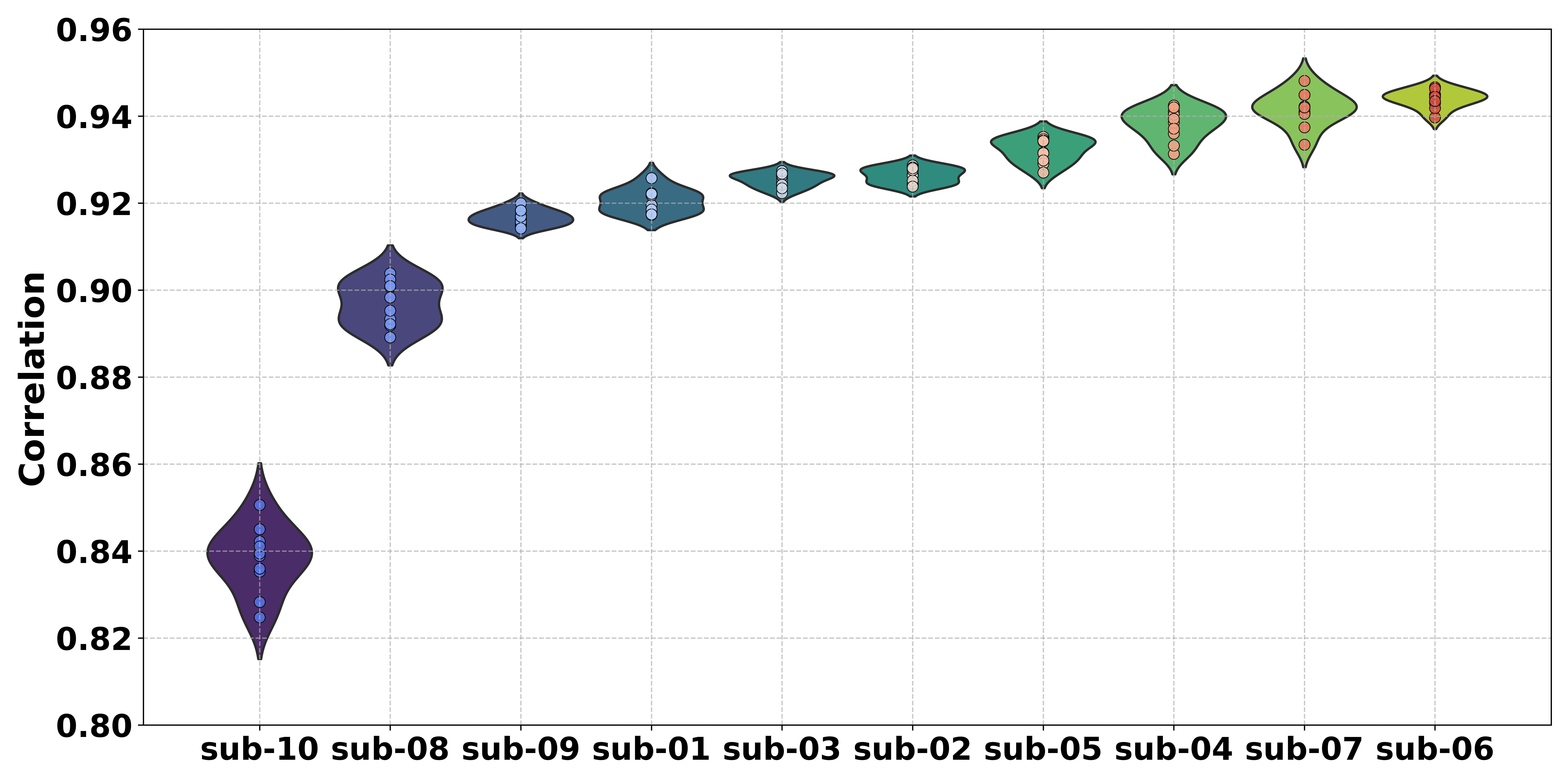}  
        \caption{Pearson correlation between predicted and original spectrograms.}  
        \label{fig:coorelations}  
        
    \end{figure}
    
\begin{figure}[t] 
        \centering 
        \footnotesize
        \includegraphics[width= \linewidth]{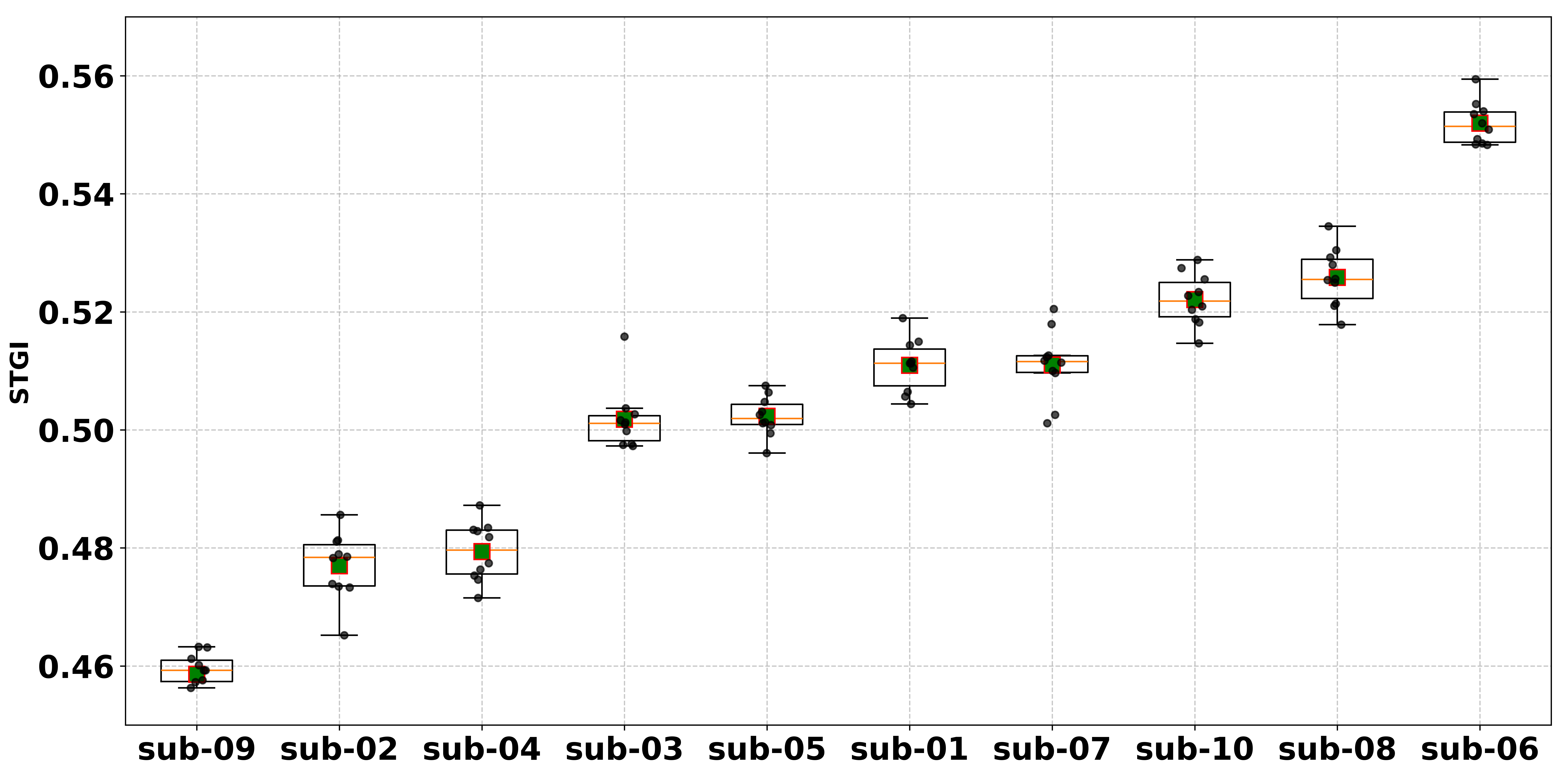}  
        \caption{STGI results between predicted and original spectrograms.}  
        \label{fig:stgis} 
    \end{figure}

    Detailed results of the distribution of each metric for the participants are shown in Figs.\ref{fig:coorelations} and \ref{fig:stgis}. In particular, Fig. \ref{fig:coorelations} depicts the distribution of the Pearson correlation coefficients computed between the original audio spectrograms and the reconstructed spectrograms in the test set. Our model demonstrates robust performance across all participants, as indicated by high correlation coefficients ranging from 0.83 in sub-10 to 0.93 in sub-06. The variation in individual correlation coefficients reflects differences in neural activity between participants, which can be attributed to the varying number of electrodes and their implantation sites. In particular, Sub-06 and Sub-07 have a higher number of electrodes implanted in the Broca and Wernicke regions of the brain compared to other participants, regions traditionally associated with cognitive processes of speech and language. In contrast, the number of implanted electrodes in Sub-10 is relatively low, particularly in the Broca and Wernicke regions. This limited electrode coverage may affect the effectiveness of monitoring or stimulating these critical areas involved in language processing.

    Fig. \ref{fig:stgis} shows a box plot diagram of the distribution of the STGI metric for the reconstructed spectrograms compared to the original audio for each participant. This metric indicates how well portions of speech signals can be understood in noisy environments. Sub-06 again is the participant with the best results, with an average STGI of 0.55, indicating better speech intelligibility, while Sub-09 has the lowest average of 0.45. The wide range of STGI values highlights differences in results between participants, which, as discussed before, could be due to variations in electrode placement, experimental conditions, or other factors.
    \subsection{Comparison with other models} \label{ssec:comparison_results}
    Table \ref{tab:comparision} presents a performance comparison of the proposed NeuroIncept Decoder model against other well-known models in the literature. The comparison is based on two metrics: PCC and STGI. The Linear Regression (LR) model, as reported by Verwoert et al. \cite{verwoert2022dataset}, achieved a PCC of 0.70, but no STGI was provided. The fully connected network (FCN) model, proposed by Band et al. \cite{band1}, obtained a PCC of 0.89 and an STGI of 0.3947. Similarly, the CNN-based model, also from Band et al. \cite{band1}, demonstrated a PCC of 0.89 and a higher STGI of 0.48. The NeuroIncept Decoder outperformed all other models, achieving the highest PCC of 0.91 and the best STGI of 0.50, indicating its superior performance in both metrics.
        
    \begin{table}[t]
        \centering
        \footnotesize
        \caption{Performance Comparison of NeuroIncept Decoder}
        \begin{tabular}{|c|c|c|}
        \hline
        \textbf{Model}  & \textbf{PCC} & \textbf{STGI} \\ \hline
        LR \cite{verwoert2022dataset} & 0.7050 & - \\ \hline
        FCN \cite{band1} & 0.8907 & 0.3947 \\ \hline
        CNN \cite{band1} & 0.8988 & 0.4839 \\ \hline
         \textbf{NeuroIncept Decoder} &  \textbf{0.9179} &  \textbf{0.5040 }\\ \hline
        
        \end{tabular}
        \label{tab:comparision}
    \end{table}

\begin{figure}[t]  
        \centering  
        \footnotesize
        \includegraphics[width= \linewidth]{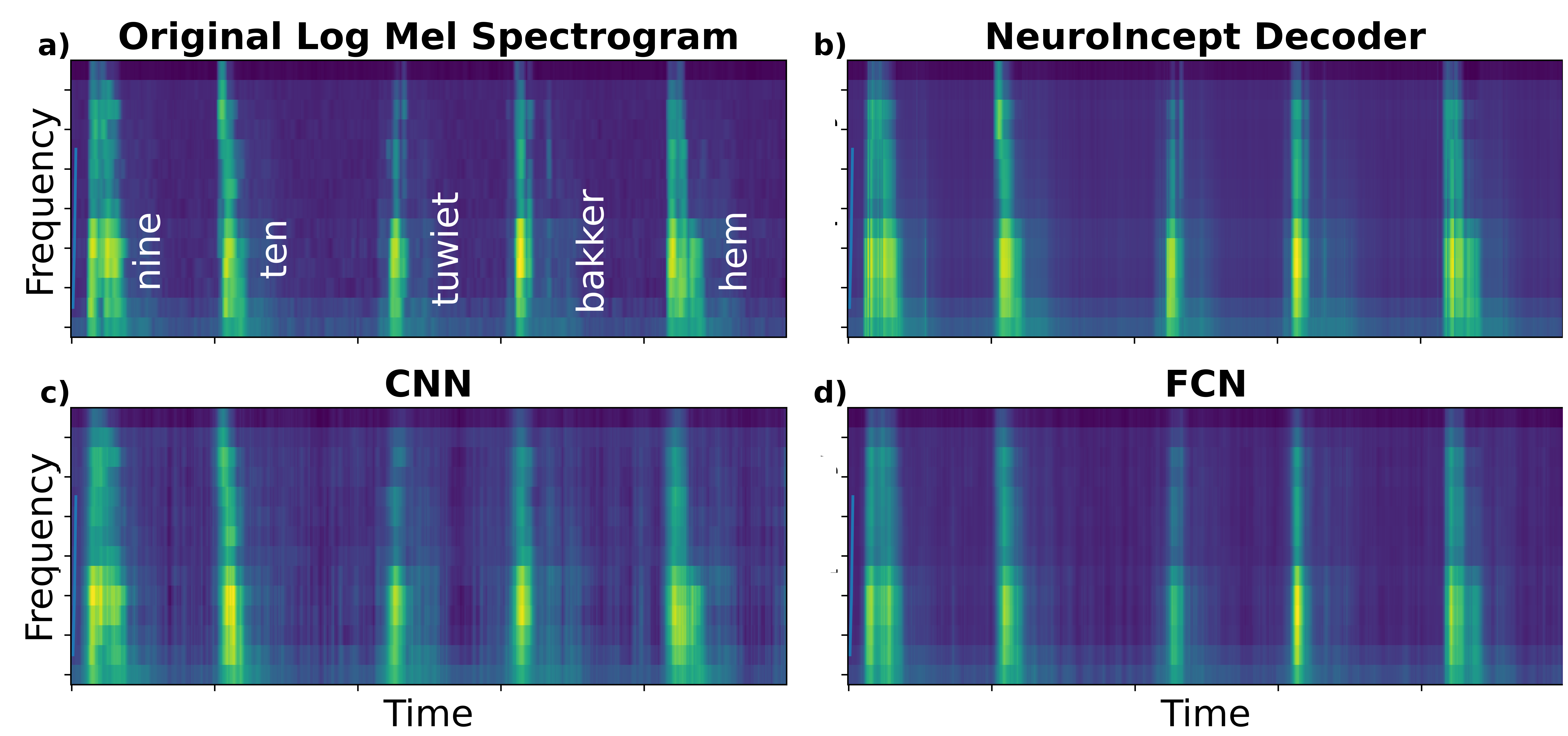}  
        \caption{Examples of logMel spectrograms for: (a) natural speech recorded by the participants, (b) speech generated by the proposed NeuroIncept model, (c) the CNN model, and (d) the FCN model. The words are same for all the plots} 
        \label{Fig:spectrogram_comparison}  
    \end{figure}

    Finally, Fig. \ref{Fig:spectrogram_comparison} shows example spectrograms produced by various models of neural activity along with the original audio recorded from participants. Our model more accurately predicts key speech characteristics, such as formants, while spectrograms generated by other techniques are smoother, suggesting lower synthesis accuracy.

\section{Conclusions}
    This study represents a significant step forward in neural decoding, demonstrating the potential to reconstruct high-quality speech directly from neural activity using advanced deep learning techniques. Our approach employs high-gamma features extracted from invasive EEG data and maps them to logMel audio features using a novel neural network architecture- the NeuroIncept Decoder. Our technique achieved high correlations between predicted and actual audio spectrograms, with values ranging from 0.83 to 0.94, highlighting its robustness in capturing key audio features. However, the modest STGI values achieved by our method indicate room for improvement in modeling the intricate temporal dynamics of speech processing in the brain.

    Future research will explore pretraining strategies for deep learning models trained on limited data, particularly approaches that utilize EEG signals linked to words or phrases without paired audio. This direction aims to enhance the system’s ability to generate real-time speech from neural data, further advancing the practical applications of neural decoding in speech restoration and brain-computer interfaces.

\vspace{12pt}

\end{document}